\pdfoutput=1  
\documentclass[acmtog,screen]{acmart}
\copyrightyear{2018}
\setcopyright{none}
\acmConference{}
\acmIsbn{}
\acmVolume{}
\acmPrice{}
\acmDOI{}
\acmYear{2018}

\renewcommand\footnotetextcopyrightpermission[1]{} 
\title{Perceptual Rasterization for Head-mounted Display Image Synthesis}

\usepackage{soul}
\usepackage{microtype}
\usepackage{amssymb}
\usepackage{amsmath}
\usepackage{booktabs}
\usepackage{hyperref}
\usepackage{multirow}
\usepackage{graphicx}
\usepackage[shortcuts]{extdash}
\usepackage[export]{adjustbox}
\usepackage{acronym}
\usepackage{gensymb}
\usepackage{siunitx}
\usepackage{placeins}
\usepackage{acronym}

\newcommand{\eg}{e.\,g.,\ }
\newcommand{\ie}{i.\,e.,\ }
\newcommand{\etal}{et~al.\ }

\newcommand{\citeetal}[1]{et~al.~\shortcite{#1}}

\newcommand{\refSec}[1]{Sec.~\ref{sec:#1}}
\newcommand{\refFig}[1]{Fig.~\ref{fig:#1}}
\newcommand{\refEq}[1]{Eq.~\ref{eq:#1}}
\newcommand{\refTbl}[1]{Tbl.~\ref{tbl:#1}}

\newcommand{\textoverline}[1]{$\overline{\mbox{#1}}$}

\def\myparagraph#1{\textit{#1.}\ }

\newcommand{\sipm}[3]{#1~$\pm$~\SI{#2}{#3}}

\def\figurePath{figures/}
\def\myfigure#1#2{\begin{figure}[h]\centering\includegraphics*[width = \linewidth]{\figurePath#1}\vspace{-.25cm}\caption{#2}\label{fig:#1}\end{figure}}

\def\mycfigure#1#2{\begin{figure*}[t]\centering\includegraphics*[clip, width = \linewidth]{\figurePath#1}\vspace{-.25cm}\caption{#2}\label{fig:#1}\end{figure*}}

\def\mysection#1#2{\section{#1}\label{sec:#2}}
\def\mysubsection#1#2{\subsection{#1}\label{sec:#2}}
\def\mysubsubsection#1#2{\subsubsection{#1}\label{sec:#2}}

\soulregister\ref7
\soulregister\cite7
\soulregister\refFig7
\soulregister\cite7
\soulregister\ref7
\soulregister\pageref7
\soulregister\shortcite7
\soulregister\eg0
\soulregister\ie0
\soulregister\etal0

\DeclareGraphicsExtensions{.png,.jpg,.pdf,.ai,.psd}
\acrodef{fov}[FOV]{Field-of-view}
\acrodef{vr}[VR]{Virtual Reality}
\acrodef{hmd}[HMD]{Head-Mounted Display}
\acrodef{crt}[CRT]{Cathode Ray Tube}
\acrodef{2afc}[2AFC]{Two-Alternative Forced Choice}
\acrodef{jnd}[JND]{Just Noticeable Difference}

\begin{document}

\author{Tobias Ritschel}
\affiliation{%
  \institution{University College London}
}
\email{t.ritschel@ucl.ac.uk}

\author{Sebastian Friston}
\affiliation{%
  \institution{University College London}
}
\email{sebastian.friston.12@ucl.ac.uk}

\author{Anthony Steed}
\affiliation{%
  \institution{University College London}
}
\email{a.steed@ucl.ac.uk}


\setcopyright{none}

\acmYear{2017}
\copyrightyear{2017}

\begin{teaserfigure}
   \includegraphics[width=\textwidth]{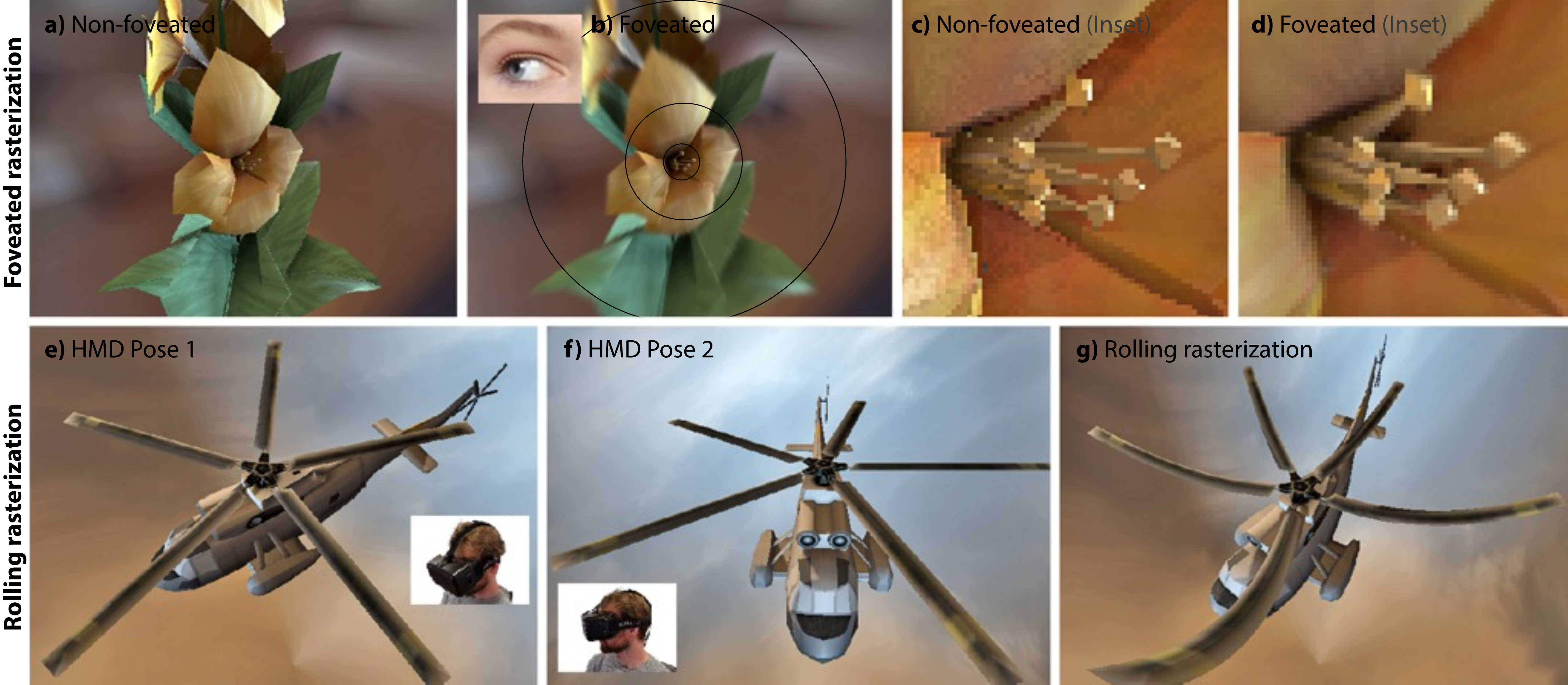}%
   \vspace{-.25cm}%
   \caption{
Perceptual rasterization is a generalization of classic rasterization to the requirements of \acp{hmd} such as foveation \emph{(top row)} and rolling image formation \emph{(bottom row)}. 
On a \ac{hmd}, most pixels appear in the periphery \textbf{(a)}.
We rasterize images with continuously-varying pixel density \textbf{(b)}.
A zoom of the the foveated area shows how a common same-shading-effort image has aliasing \textbf{(c)}, while our result benefits from higher pixel density, resulting in super-sampling \textbf{(d)}. 
In common rasterization, each pixel on the display is effectively sampled at the same simulation time ($t=0$ for the first frame \textbf{(e)} and $t=1$ for the next frame \textbf{(f)}).
When displayed on a ``rolling'' HMD display, where pixels are illuminated at different points in time, latency is introduced: the rightmost pixel is outdated by ca.~16\,ms.
Our rolling rasterization \textbf{(g)} allows spatially-varying time: starting at $t=0$ on the left of the image and increasing to $1$ on the right.
}
   \label{fig:Teaser}
\end{teaserfigure}

\begin{abstract}
We suggest a rasterization pipeline tailored towards the need of \acp{hmd}, where latency and field-of-view requirements pose new challenges beyond those of traditional desktop displays.
Instead of rendering and warping for low latency, or using multiple passes for foveation, we show how both can be produced directly in a single perceptual rasterization pass.
We do this with per-fragment ray-casting. This is enabled by derivations of tight space-time-fovea pixel bounds, introducing just enough flexibility for requisite geometric tests, but retaining most of the the simplicity and efficiency of the traditional rasterizaton pipeline.
To produce foveated images, we rasterize to an image with spatially varying pixel density.
To reduce latency, we extend the image formation model to directly produce ``rolling'' images where the time at each pixel depends on its display location.
Our approach overcomes limitations of warping with respect to disocclusions, object motion and view-dependent shading, as well as geometric aliasing artifacts in other foveated rendering techniques.
A set of perceptual user studies demonstrates the efficacy of our approach.
\end{abstract}

\maketitle

\thispagestyle{empty}

\mysection{Introduction}{Introduction}
The use cases of HMDs have requirements beyond those of typical desktop display-based systems.
Completely subsuming the user's vision, the HMD and system driving it must maintain low and predictable latency to facilitate a sense of agency and avoid serious negative consequences such as breaks-in-presence~\cite{Slater2002}, simulator sickness~\cite{Buker2012}, and reduced performance~\cite{Ellis1999a}.
This challenge is exacerbated by other characteristics of \acp{hmd}, such as high \ac{fov} and resolution.
Further, as human vision has varying spatial resolution with a rapid fall-off in the periphery, much of this computational effort is wasted.

Ray-tracing could cast more rays to the foveal area (foveation) and update the view parameters during image generation (low latency).
Regrettably, ray-tracing remains too slow in large and dynamic scenes.
Traditional rasterization efficiently draws an image, but with uniform detail. It does not take advantage of how that image will be perceived.
Here, we suggest \emph{perceptual rasterization} that retains most of the efficiency of rasterization, but has additional optimizations that are especially beneficial for \acp{hmd}: low-latency and foveation.

This is achieved by generalizing common OpenGL-style rasterization.
Our foveated rasterization can work with \acp{hmd} that provide eye-tracking data, such as the FOVE~\shortcite{fove2018}, allowing  rasterization into a framebuffer with a non-constant pixel density that peaks at the fovea.
Our rolling rasterization gives every column of pixels a different time and can be used on \acp{hmd} with rolling displays, such as the Oculus Rift DK2, that illuminate different spatial locations at different times.
The techniques can be used together.

After discussing previous work (\refSec{PreviousWork}), we will, describe our novel perceptual rasterization pipeline (\refSec{PerceptualRasterization}) before presenting the specific time, space and retinal bounds in \refSec{Bounds}. In \refSec{Results} we present image results and analysis and in \refSec{UserStudies} we present four user studies that demonstrate the efficacy of perceptual rasterization. 

\mysection{Previous Work}{PreviousWork}

\myparagraph{Foveated rendering}
The wide \acp{fov} ($100$ degrees and more) found in current \acp{hmd} \cite{toth2016comparison,patney2016towards,weier2017perception,fove2018} require higher resolutions and therefore increasing amounts of memory and bandwidth on the GPU.
At the same time, only a small percentage of the screen falls onto the fovea, where the highest resolution is required.
This makes foveated rendering particularly important for \acp{hmd}.
In-HMD eye tracking \cite{stengel2015affordable,fove2018} is required to know the fovea's location.

Guenter~\etal\shortcite{guenter2012foveated} demonstrate a working end-to-end foveated system based on rasterization.
To achieve foveation, they rasterize in multiple passes (three in their example) to individual images with different but uniform pixel densities.
We also use rasterization, but into an image with continuously varying pixel density and in a single pass.
The work of Patney~\etal\shortcite{patney2016towards} applies blur and contrast enhancement to the periphery to hide artifacts.
In doing so, they can further reduce the size of the highest resolution foveal region without becoming noticeable.
Reducing shading in the periphery is discussed by He~\etal \shortcite{he2014extending}.
However, this does not increase pixel density in the fovea, whereas our approach provides substantial super-sampling of both shading and geometry. 
%
%

\myparagraph{Display latency}
In \ac{vr} systems to date, an important delay that contributes to the end-to-end latency is the interval $[t_\mathrm s, t_\mathrm e]$ during which a pixel will be displayed.
The longer the interval, the more ``outdated'' a stimulus will become: if each pixel holds a constant value for 1/60 of a second, at the end of the interval $t_\mathrm e$ the image may deviate significantly from the ideal representation of the state of the virtual world at the time it was rendered (at or before $t_\mathrm s$).
In combination with head or eye motion, this leads to hold-type blur \cite{sluyterman2006needed,didyk2010perceptually}.

\myfigure{Capture}{
\textbf{a)} Seven frames (24\,ms) high-speed capture (Casio Exilim EX-ZR1000) of an HDK 2 HMD (twin) display.
Specific locations are illuminated \emph{(blue)} at specific points in time.
\textbf{b)} Time-varying illumination of a \SI{4}{\milli\metre} band of an Oculus DK2 display captured with a photodiode and a PicoScope 6402B.
}

To compensate for these negative effects, designers use displays with increasing refresh rates, and lower persistence.
Increased refresh rates reduce apparent latency by limiting the maximum age of a given pixel.
Low persistence displays illuminate the screen for a time far below the refresh period of the display.
This reduces artifacts such as blur.
Some of these low persistence displays use a ``global scan'', in which the entire display is illuminated at once.
These have two complications: the display is much darker and global changes in brightness can produce noticeable flicker.
Low brightness is a relatively minor issue for HMDs because the user's vision can adapt.
However flicker will be very noticeable, as the human ability to detect flicker is stronger if the target is large (the Granit-Harper~\shortcite{granit1930comparative} law).
An alternative low persistence display technology behaves similarly to traditional cathode ray tubes.
That is, pixels are illuminated for a short period as they are updated.
We consider such displays to have a ``rolling scan'' (\refFig{Capture}).
Drawbacks and benefits of such a display are discussed by Sluyterman~\shortcite{sluyterman2006needed}.
They exhibit less flicker (as the target is smaller \cite{granit1930comparative}) while remaining resistant to blur.
Both global and rolling scan displays will show outdated stimuli, as there is still a delay between the time $t$ a pixel is rendered, and $t_\mathrm s$ when it is displayed.

Our solution is to produce a \emph{rolling} image, where pixels at different spatial locations correspond to different points in time \cite{friston2016construction}.
This is analogous to a rolling shutter sensor which captures light at different points in time for different sensor locations.

\myparagraph{Ray-tracing}
Both rolling and foveated images can be generated by ray-tracing: rays are free to use a different time value to intersect the virtual world and more rays could be sent to the fovea \cite{stengel2016adaptive,weier2016foveated}.
Low-latency ray-tracing has been demonstrated at interactive rates for simple scenes with specialized hardware~\cite{friston2016construction}.
Foveated ray-tracing is demonstrated by
Stengel~\etal\shortcite{stengel2016adaptive} in a system that adaptively sends more rays into perceptually important areas, including the fovea.
Weier~\citeetal{weier2016foveated} also describe a solution that provides foveated ray-tracing for \acp{hmd} in real-time.
Both systems require scenes that fit the assumptions of interactive ray-tracing.

Significant advances in ray-tracing have been made \cite{wald2014embree}, but it is still typically considered too slow for modern interactive applications with complex dynamic scenes, such as computer games.
It is also not clear how a modern ray tracer making use of bounding volume hierarchies (BVH) would handle a more continuous approximation of frame time.
Rebuilding the BVH for every pixel would certainly be less than ideal. 

\myparagraph{Warping}
One source of latency is the time expended between beginning a render and displaying it. One way to counteract this is to warp, \ie deform, the final image, accounting for changes in viewpoint during the render. Early approaches changed which regions of an image were read out \cite{regan1994priority,OculusSDK}, or drew points \cite{chen1993view} or grids \cite{mark1997post}. 
Modern approaches such as \emph{Asynchronous Time Warping} (ATW) \cite{antonow2015} incorporate a number of these techniques to compensate for multiple sources of latency.
The main drawback of warping is that it suffers disocclusion artefacts. Some techniques can help ameliorate these, such as perceptually improved hole filling \cite{didyk2010perceptually,schollmeyer2017efficient}. Alternatively the result can be improved by changing the images provided to the algorithm itself~\cite{Reinert2016ProxyIBR}.
No deformation however can reveal what is behind a surface. 
Our images have no disocclusion artefacts, and also support correct specular shading.

\myparagraph{Shading latency}
Due to latency, specular shading is also incorrect as highlights depend on the moving viewpoint that is frozen at the start of the frame in classic pipelines \cite{antonow2015}.
This could be resolved by ray-tracing, but would still produce problems if combined with warping.
Perceptual rasterization correctly resolves specular shading.

\myparagraph{Non-standard rasterzation}
A simple solution to achieve both rolling and foveated images is to change the vertex shader \cite{brosz2007single} from a linear to a non-linear projection, such as first done for shadow mapping \cite{brabec2002shadow}.
Doing this for latency compensation or foveation results in holes, in particular if primitives are large or close to the camera, as primitive edges remain straight \cite{brosz2007single}.
Our approach is a type of non-linear rasterization \cite{gascuel2008fast}. 
Toth~\etal\shortcite{toth2016comparison} suggest single-pass rendering into spatially neighboring but linear sub-projections to address the non-uniform pixel distribution in \acp{hmd}, but do not account for eye tracking.
Rasterization has been made more flexible in stochastic rasterization \cite{akenine2007stochastic,brunhaver201hardware,mcguire2010real}, but we are not aware of an approach to produce rolling or foveated images directly using rasterization in a single pass.
In particular, we derive non-trivial bounds specific to our projection that drastically improve the sample test efficiency, \ie how many fragments need to be tested against each primitive \cite{pineda1988parallel,laine2011clipless}.

\mysection{Perceptual Rasterization}{PerceptualRasterization}
We first describe the general perceptual rasterization pipeline before deriving specific bounds enabling its application to foveation, rolling and both.
The key is to achieve just enough ray tracing-like flexibility while retaining the efficiency of rasterization.

Let us first recall rasterization and ray-tracing: ray-tracing iterates over pixels and finds the primitive mapping to them, while rasterization iterates over primitives and maps them to pixels.
Our technique is a hybrid of these approaches. To decide what pixels a primitive maps to, the rasterisation essentially performs ray-primitive intersections \cite{pineda1988parallel} followed by a $z$-test. A correct, but slow, solution would be to test all primitives against all pixels.
Instead, the approach becomes fast by using tight \emph{primitive-pixel bounds}: ideally, a compact, easy-to-compute subset of pixels is found for the projection of each primitive in a first step, and only the rays going through these pixels are tested against the primitive.

The idea of perceptual rasterization is to construct such pixel-primitive bounds for the requirements of \acp{hmd}.
To this end, we will next propose different \emph{ray-primitive models} we use (\refSec{RayPrimitiveModels}), before describing the pipeline in detail in \refSec{Pipeline}.
The actual bounds are then derived in \refSec{Bounds}.

\mysubsection{Ray-primitive Models}{RayPrimitiveModels}
The interaction between rays and primitives required on an \ac{hmd} are not arbitrary, as, say, in path tracing, but have a very specific layout in time, space and the retina, which we will later exploit to construct appropriate bounds.
We will now discuss the ray-primitive models required for common, as well as our foveated, rolling and jointly foveated-rolling rasterization.


\mysubsubsection{Foveated}{FoveatedModel}
To retain the simplicity of rasterization on a regular grid, we seek inspiration from information visualization \cite{furnas1986generalized} and directly from cortical magnification theory \cite{daniel1961representation}: to give more importance to an area, it simply needs to be magnified.
So instead of increasing the pixel density in the fovea, we just magnify it.

\myfigure{FoveationMapping}{
Foveation and unfoveation function \emph{(a)} and domains \emph{(b)}.}

\mycfigure{Overview}{
Overview of perceptual rasterization.
Common rasterization \textbf{(a)} produces images at a fixed time and uniform pixel density.
We suggest to account for primitive motion, here shown as two frames \textbf{(b)} and non-uniform pixel density, here visualized as iso-lines (\textbf{c}). 
Primitive-ray interaction is bound, here using a rectangle \textbf{(d)} and intersected \textbf{(e)} to produce a rolling and foveated image to be shaded (\textbf{f}).
(Depiction uses a monoscopic HMD display for simplicity.)
}

\myparagraph{Domain}
We suggest an image domain where the ray (or pixel) density depends on a function $p(d)\in(0,\sqrt 2)\rightarrow\mathbb R^+$, where $d$ is the distance to the foveation point $\mathbf x_\mathrm f$.
In common rasterization, this function is a constant: $1$ (\refFig{FoveationMapping} a, constant line).
For foveated rendering, it is higher close to the fovea ($d$ is small) and lower than $1$ for the periphery ($d$ is large) (\refFig{FoveationMapping}, a, yellow line).

$p$ can be any foveation function, whether physiologically based \cite{daniel1961representation} or empirically based \cite{weier2017perception,patney2016towards}.
The size of the foveated region, and therefore $p$, must account for non-idealities such as imperfect tracking and suboptimal frame rates. These may also change over time. Therefore we refrain from using any analytic model and instead assume that the function is arbitrary, subject to the constraints below, and free to change every frame.

Given $p$, we define another function $q(\mathbf x)\in(-1,1)^2\rightarrow(-1,1)^2 : \mathbf x_\mathrm f+ norm(\mathbf x-\mathbf x_\mathrm f) \cdot p(||\mathbf x-\mathbf x_\mathrm f||)$. This function essentially scales $\mathbf x$ by $p$, away from the gaze position. Near the center, this results in stretching, as the pixel density is larger than 1. In the periphery, compression, as fewer pixels are required (\refFig{FoveationMapping}, b). We also define $q^{-1}$, to be $q$ but with $p^{-1}$ in place of $p$. $p^{-1}$ is the inverse of $p$. Note that $d$ is not a scaling factor but an exact distance. Thus $p$ maps an unfoveated distance to a foveated distance, and $p^{-1}$ maps it back. $q$ and $q^{-1}$ use these functions to do the same for pixel locations. We refer to these pixel transformations as to ''foveate" and ''unfoveate".
This necessitates that $p$ is invertible. Any monotonic $p$ can be inverted numerically in a pre-processing pass, if an analytic inversion is non-trivial.

\myparagraph{Display}
After rasterizing all primitives, the foveated image $I_\mathrm f$ has to be converted back into an unfoveated $I_\mathrm u$ one for display.
This imposes several challenges for filtering: $q^{-1}$ is heavily minifying in the center and heavily magnifying in the periphery.
A simple and fast solution is to create a MIP map for the foveated image and then evaluate $I_\mathrm u(\mathbf x)=I_\mathrm f(q^{-1}(\mathbf x))$ using proper tri-linear MIP mapping and a 3-tap cubic filter (0.6\,ms in 1024$\times$1024 on an Nvidia GTX~980~GPU).
A higher-quality version (1.6\,ms in 1024$\times$1024, same GPU) computes \[
L_\mathrm d(\mathbf x) = 
\sum_{y\in 5\times 5} 
L_\mathrm c(q(\mathbf x) + \mathbf y)
\cdot
r(||\mathbf x - q^{-1}(q(\mathbf x) + \mathbf y))||)
,
\]
where $L_\mathrm d$ is the display image,
$L_\mathrm c$ the foveated imaged, and $r$ an arbitrary, \eg Gaussian, reconstruction filter parametrized by distances in the display image domain.
Such an operation effectively computes the (irregular-shaped) projection of the display's reconstruction filter into the cortical domain.


\mysubsubsection{Rolling}{RollingModel}
Here, the ray direction and position at a certain pixel depends on the time that pixel is displayed.
When testing a ray through a given pixel, the state of the primitive intersected also has to be its state at the time the pixel is displayed.

\myparagraph{Display}
We consider a rolling-scan display to have three properties: rolling illumination, a short hold-time, and we must be able to predict the absolute head pose at any point in the interval [$t_\mathrm s$,$t_\mathrm e$].

First, a \emph{rolling scan} implies that different parts of the display are visible at different times.
The term ``rolling'' is chosen as an analogy to a camera's rolling shutter sensor.
A classic \ac{crt} is an example of a rolling scan display.
Most LCDs these days perform a global synchronized illumination of all pixels at once.
OLEDs, such as those used in the DK2 and other \acp{hmd} sometimes use rolling illumination.

We will formalize this as a \emph{rolling}-function
$
r(\mathbf x)
\in(0,1)^2
\rightarrow
(0,1):
\mathbf x
\cdot
\mathbf d
$
that maps a (unit) spatial location $\mathbf x$ to a (unit) point in time at which the display will actually show it by means of a skew direction $\mathbf d$.
$\mathbf d$ depends on the properties of an individual display.
For example $\mathbf d=(0,.9)$ describes a display with a horizontal scanout in the direction of the x-axis and a (blank) sync period of 10\,\% of the frame period. For the DK2, $\mathbf d=(1,0)$ based on behavior profiled with an oscilloscope (\refFig{Capture}).

Second, the display has to be low persistence (\emph{non-hold-type}), \ie a pixel is visible for only a short time relative to the total refresh period.
A \ac{crt} is typically of this type. \ac{crt} phosphor has a decay that typically reduces brightness by a factor of 100 within one millisecond (Fig.~1 in  \cite{sluyterman2006needed}).

Third, we assume that the model-view transformation can be linearly interpolated across the animation interval and that vertices move along linear paths during that time.
More general motion is possible, but not for the tightest bound (Zenon's bound), which uses an analytic derivation requiring linearity.

\mysubsubsection{Joint foveated-rolling}{JointModel}
The composition $r \circ q(\mathbf x)$ of the above.

\mysubsection{Pipeline}{Pipeline}
An overview of perceptual rasterization is see in \refFig{Overview}, d--f.
We extend a classic OpenGL-style rasterization pipeline using vertex, geometry and fragment programs (VP, GP and FP) to produce a typical deferred shading buffer from primitives in two steps: \emph{bounding} and \emph{intersecting}.
We will explain how to bound tightly and efficiently for the different models later in \refSec{Bounds}.

\myparagraph{Bounding}
Input to the VP are the world-space vertex positions $v_\mathrm s$ at the beginning and $v_\mathrm e$ at the end of the frame interval.
Additionally, the VP is provided two model-view-projection matrices $\mathsf M_\mathrm s$ and $\mathsf M_\mathrm e$ that hold the model and view matrices at the beginning and the end of the frame interval.
The VP transforms both the start and the end vertex, each with the start and the end matrix ($\mathsf M_\mathrm s v_\mathrm s$ and $\mathsf M_\mathrm e v_\mathrm e$), and passes this information on to the GP.
Note, that no projection is required at this step.

Input to the GP is the tuple of animated camera-space vertices $S=(
v_{\mathrm s, 0},
v_{\mathrm e, 0},
v_{\mathrm s, 1},
v_{\mathrm e, 1},
v_{\mathrm s, 2},
v_{\mathrm e, 2}
)$, \ie an animated camera space triangle.
The GP \emph{bounds} the projection of this space-time triangle with a 2D primitive, such that all pixels that would at any point in time be affected by the triangle are covered by the new bounding primitive $B$.
The geometry program passes the space-time triangle on to the fragment program as (flat) attributes.
Note, that the bounding primitive $B$ is not passed on from the GP to the FP: It is only required as a proxy to determine the pixels to test directly against $S$ (and not $B$) \ie what pixels to rasterize.
The fragment program then performs the intersection test described next.

\myparagraph{Intersection}
The fragment program is now executed for every pixel $i$ that could be affected by the primitive's bound.
Note that this test is the same regardless of what bounding is used.
To decide if the pixel $\mathbf x_i$ actually is affected by the space-time triangle, we intersect the ray $R_i$ at this pixel with the triangle at time $r(\mathbf x_i)$.
The entire triangle, its normals, texture coordinates and material information, were emitted as \texttt{flat} attributes from the GP.
Note, that $R$ depends on the time as well: every pixel $i$ has to ray-trace the scene at a different time following $r$.
For foveation, $R_i$ is not formed by a pin-hole model but follows $q$.
The joint model distributes rays according to $r\circ q$.
The position of the entire triangle at time $r(\mathbf x_i)$ is easily found by linear interpolation of the vertex motion.
This results in a camera-space triangle $T_i$, that can be intersected with $R_i$ using a 3D ray-triangle intersection test.
If the test fails, nothing happens.
If the test passes, the fragment is written with the actual $z$ value of the intersection and with common $z$ buffering enabled.
This will resolve the correct (\ie nearest to the viewer) fragment information.
For every pixel there is a unique time and fovea location, and hence distances of multiple primitives mapping to that pixel are $z$-comparable.
This is key to make perceptual rasterization possible when primitives are submitted in a streaming fashion in an arbitrary order. 

\myparagraph{Shading}
Shading has to respect the ray-primitive model as well: the time at every pixel is different for the rolling and joint model, having the implication that parameters used for shading, such as light and eye position should also be rolling and differ per pixel.
This again can be done by simple linear interpolation.
Note that shading is not affected by foveation.


\mysection{Bounds}{Bounds}
A key technical contribution of this paper is the derivation of tight and efficiently computable bounds for the ray-primitive model required for modern \acp{hmd}.

\mysubsection{Foveation bounds}{FoveationBounds}

\myfigure{FoveatedBounding}{
Foveated bounding.
\textbf{a)} the original primitive.
\textbf{b)} the foveated primitive.
\textbf{c)} the simple bounds displaces the original edges.
\textbf{d)} the advanced bound first foveates the edges and then bounds the displacement.
}

As there is no closed-form foveation function available, we cannot derive a closed-form solution as will be done for rolling rasterization.
This might become possible in future work when using an analytic foveation function. For now, the monotonicity property still allows for tight bounds that are quick to compute (\refFig{FoveatedBounding}).

The key is to use $q$ and $q^{-1}$.
The bounding geometry we generate will always consist of a convex polygon with six vertices, and does not require a convex hull computation.
Every even pair of vertices is produced by bounding a single edge of the original triangle.
Every odd pair joins the start and end of a bounding edge produced from a primitive edge.
The remaining task is then to bound a single triangle edge from $\mathbf x_0$ to $\mathbf x_1$.
We have derived two bounds, a simple and a tighter recursive bound.

\mysubsubsection{Simple}{SimpleFoveationBound}
Here, the bounding edge is assumed to be parallel to the original edge (\refFig{FoveatedBounding},c)
All we need to find is the maximal positive distance along the normal from the edge joining $\mathbf x_0$ and $\mathbf x_1$ 
\begin{multline*}
\Delta_\mathrm{max}=
\max_{s\in(0,1)}
\{
\Delta(s) = (\eta_ s(s)-\eta_ c(s))
\cdot
n(\eta_ s(0), \eta_ s(1))
\}\\
\eta_\mathrm s(s)=\mathbf x_0+s(\mathbf x_1-\mathbf x_0)
\quad
\text{and}
\quad
\eta_\mathrm c(s)=q(\mathbf x_0+s(\mathbf x_1-\mathbf x_0)),
\end{multline*}
where $n$ creates a direction orthogonal to the line between its two arguments.
As the distance is a convex function, it can be minimized using a ternary search that converges to a pixel-precise result in $\log(n)$ steps, if $n$ is the number of possible values, here, the number of pixels on the edge.
Consequently, for a 4\,k image example, bounding requires $3\times2\times\log(4096)=96$ multiply-adds and dot products per triangle at most, but typically much less as triangle edges are shorter.

\mysubsubsection{Recursive}{RecursiveFoveationBound}
Consider the original vertices $\mathbf x_0$ and $\mathbf x_1$ (\refFig{FoveatedBounding}, a) and the foveation $q(\mathbf x_0)$ and $q(\mathbf x_1)$ of these vertices (\refFig{FoveatedBounding}, b).
While the simple bound displaces relative to the straight original edge from $\mathbf x_0$ to $\mathbf x_1$ (\refFig{FoveatedBounding}, c) the new recursive bound will displace relative to the straight edge $q(\mathbf x_0)$ to $q(\mathbf x_1)$ (\refFig{FoveatedBounding}, d):
\[
\eta_\mathrm s(s)=q(\mathbf x_0)+s(q(\mathbf x_1)-q(\mathbf x_0))
\]
This is possible, as the edge has to be straight, but not necessarily the ``original'' one.
The resulting bound is tighter, \ie the blue area is smaller than the yellow one in \refFig{FoveatedBounding}.
Note, that the normal for a different straight edge is also different, as $q$ is a nonlinear function: an edge joining a point close to the origin and a point farther from the origin will change its slope as both are scaled differently.

\mysubsection{Rolling bounds}{RollingBounds}

\myfigure{Bounds}{
Rasterization bounds for a space-time triangle moving across the screen.
The triangle starts at a position where the frame time already is $.3$ and ends where frame time is $.6$.
Consequently, it can not cover the full convex hull, but only the convex hull of a spatio-temporal subset.
We identify this region, resulting in an increased sample test efficiency (cf.\ the ratio of areas of ``Hull'' and ``Adaptive'').
Finally, there is an analytic solution to when exactly the rolling beam will catch up with a moving primitive allowing for even tighter bounds (``Zenon'').
}

\mysubsubsection{Boxes}{BoxBound}
A reasonably tight bound for the time-space triangle $S$ as defined in \refSec{Pipeline}, is the 2D bounding box
\[
B=\mathtt{bbox}
\{
\mathcal P(S_{i,j}, t)
|
i\in\{\mathrm s, \mathrm e\},
j\in\{0,1,2\},
t\in\{0,1\}
\}
\]
of all vertices in the start and end of the frame, where
$\mathtt{bbox}$ builds the 2D bounding box of a set of points and
$\mathcal P$ is the projection of a point at time $t$, \ie multiplication with a time-varying matrix followed by a homogeneous division (``Quad'' in \refFig{Bounds}).

\mysubsubsection{Convex hull}{ConvexHullBound}
A bounding box would create substantial overdraw for thin and diagonal primitives.
Investing time to produce tighter bounding primitives can be worthwhile as it reduces the amount of work done for each pixel (``Hull'' in \refFig{Bounds}).
Fortunately, all points of a triangle under linear motion fall into the convex hull of its vertices \cite{akenine2007stochastic}.
We can therefore replace the operator $\mathtt{bbox}$ by the convex hull of a set $\mathtt{hull}$ that could be implemented efficiently \cite{mcguire2010real} (our current implementation uses a GLSL quick hull implementation).
For primitives intersecting the near plane we proceed as similar to McGuire~\etal\shortcite{mcguire2010real}: all primitives completely outside the frustum are culled; primitives completely in front of the camera (but maybe not in the frustum) are kept, and those that intersect the near plane are split by this plane and their convex hull is used.
We found using a convex hull of up to 15 points (there are 15 edges between 6 space-time vertices \cite{mcguire2010real}) resulted in higher overall performance than when using the simpler bounding box.

\mysubsubsection{Adaptive}{TimeSpaceBound}
While convex hulls are tight spatially, the rolling case allows for a surprisingly tighter bound under some simple and reasonable assumptions on $w$, the mapping from pixel locations to frame times (``Adaptive'' in \refFig{Bounds}).
The key observation is that a rolling space-time triangle only has to cover 
\[
B=\mathtt{hull}
\{
\mathcal P(S_{i,j}, t)
|
i\in\{\mathrm s, \mathrm e\},
j\in\{0,1,2\},
t\in\{t_\mathrm{min}, t_\mathrm{max}\}
\},
\]
where the triangle-specific time interval $(t_\mathrm{min},t_\mathrm{max})$ is found by mapping back 2D position to time
\[
t_\mathrm{min} = 
\min
\{
w^{-1}
\mathcal P(S_{i,j}, t)
|
i\in\{\mathrm s, \mathrm e\},
j\in\{0,1,2\},
t\in\{0,1\}
\}.
\]
The maximal time $t_\mathrm{max}$ is defined by replacing the minimum with a maximum operation.
In other words, to bound, we first project all six vertices with time 0 and 1, to get bounds in 2D but then find the maximal and minimal time at which these pixels would be relevant.
As this time span is usually shorter than the frame \ie $t_\mathrm{min}\gg t_\mathrm s$ and $t_\mathrm{max}\ll t_\mathrm e$, the spatial bounds also get tighter.

\mysubsubsection{Zenon's hull}{ZenonsBound}
The problem of bounding where the rolling scan will ``catch up'' with the projection of a moving triangle has similarity with Zenon's paradoxon where Achilles tries to catch up with the tortoise \cite{aristotle1929physics} (\refFig{Zenon}, a).

\myfigure{Zenon}{Linear (\textbf{a}) and perspective (\textbf{b}) Zenon's paradoxon (see text below).}

If Achilles starts at $x_\mathrm s$ and moves at constant speed $\dot x_\mathrm s$, it will reach (other than what the paradoxon claims) a tortoise at position $x_\mathrm p$ with 1D speed $\dot x_\mathrm p$ at the time $t$ where
\[
x_\mathrm s+t\dot x_\mathrm s=
x_\mathrm p+t\dot x_\mathrm p,
\text{ \quad which occurs at \quad }
t
=
\frac{x_\mathrm s-x_\mathrm p}
{\dot x_\mathrm s-\dot x_\mathrm p}.
\]
The same holds for a rolling scan (Achilles) catching up with a vertex (tortoise).
Regrettably, in our case, the rolling scan moves in image space, while the primitive moves in a 2D projective space (horizontal $x$ component and projective coordinate $w$) from spatial position $x$ with speed $\dot x$ and projective position $w$ with speed $\dot w$ (\refFig{Zenon}, b).
This can be stated as
\[
x_\mathrm s+t\dot x_\mathrm s=
\frac
{x_\mathrm p+t\dot x_\mathrm p}
{w_\mathrm p+t\dot w_\mathrm p},
\]
which is a rational polynomial with a unique positive solution
\begin{equation}
t = 
-\frac
{
(\sqrt{4 x_\mathrm s \dot w_\mathrm p + \dot x_\mathrm s^2 - 2 \dot x_\mathrm s w_\mathrm p + w_\mathrm p^2} - \dot x_\mathrm s + w_\mathrm p)
}{
2 \dot w_\mathrm p
}
.
\label{eq:Zenon}
\end{equation}
To produce the final bounds, the time $t_i$, and the 2D position $x_i$ at this time, is computed for each of the six vertices of the space-time triangle.
The convex hull of the $x_i$ is the final bounding geometry.

\mysubsection{Joint Foveated-rolling bounds}{JointRasterization}
A joint approach for rolling and foveation operates similarly to the foveation-only approach.
To add rolling to foveation, we add the rolling transformation to $q$ (\refFig{CombinedRasterization}).
Order is important: the rolling time coordinate has to depend on where the pixel will effectively be displayed in the non-foveated domain.
Let $\mathbf x_0$ and $\mathbf x_1$ be the original world coordinates of that edge
The new edge functions are therefore \[
\eta_\mathrm s(s)=
\mathcal Q(\mathbf x_0)+s(\mathcal Q(\mathbf x_1)-\mathcal Q(\mathbf x_0))
\quad
\text{and}
\quad
\eta_\mathrm c(s)=
\mathcal Q(\mathbf x_0+s(\mathbf x_1-\mathbf x_0))
\] where
$\mathcal Q$ is the joint action of rolling and foveation $\mathcal Q(\mathbf x)\in\mathbb R^3\rightarrow\mathbb R^2:\mathcal Q(\mathbf x)=q(\mathcal P(\mathbf x, t))$.
The time $t$ can be found using \refEq{Zenon}.

\myfigure{CombinedRasterization}{
Joint rolling-foveated rasterization.
\emph{a)} One original edge of a primitive in orange.
\emph{b)} rolling of the same edge results in the blue curve.
\emph{c)} Foveation of that curve leads to another pink curve, that is bound from the line joining its ends, adding the gray area.
}

\mycfigure{FoveatedResults}{
Foveation results.
The first column shows the result we produce, fovea marked in yellow.
The second to fourth columns shows the foveated region using non-foveated rendering, our approach, and a $4\times 4$ super-sampling reference.
Quantitative evaluation is found in \protect\refTbl{FoveatedMain}.
}

\mycfigure{RollingResults}{Results of our rolling rastrization approach.
Different rows show different scenes.
The first column shows the input image.
The result of warping is shown in the second, where disocclusions were filled with gray.
The third column shows our approach.
The fourth and fifth columns shown the inset areas from columns two and three.
Quantitative evaluation is found in \protect\refTbl{RollingMain}.
Please, see the supplemental video for animated versions of these results.
}

\mysection{Results}{Results}
We discuss qualitative (\refSec{QualitativeResults}) and quantitative (\refSec{QuantitativeResults}) results.

\mysubsection{Qualitative}{QualitativeResults}

\myparagraph{Foveation}
Results of our foveated rasterization approach are seen in \refFig{FoveatedResults}.
Our image was produced by foveating the center using a simple power-falloff $p(x)=x^2$ foveation function.
The inset shows a $32\times 32$ patch. The reference was produced by $4\times 4$ super-sampling.

We see that the amount of detail varies across the image in the first column.
While the center is sharp, yet super-sampled, the periphery has less detail, yet blurred with a high-quality cubic filter.
In the common condition (second column) the fine hairs of the hairball lead to almost random results without super-sampling, while our result remains smooth and similar to the reference.
The same is true for the fine geometric details in the car's grill.
In the \textsc{Children} scene, the super-sampling of shading is salient.

The common images were produced using the same memory, the same shading effort and not less than half the compute time than ours (third column), yet the differences are visible.
At the same time, the reference (fourth column), uses 16 times more memory and shading effort and is more than twice the compute time than ours, yet the differences are subtle.

\myparagraph{Rolling}
Images produced by our rolling rasterization approach can be seen in \refFig{RollingResults}.
A non-rolling image is seen in the first column.
The second and third columns contain rolling images where the camera has both translated and rotated during a rolling scan-out from left to right.
The second column shows image warping using a pixel-sized grid, where triangles that have a stretch that differs by more than a threshold are culled entirely \cite{mark1997post}.
Disoccluded areas are marked with a checkerboard pattern.
The third column shows the results produced by our approach.
The fourth and fifth columns show insets from the second and third row.
The scenes were intentionally chosen to contain large polygons, which are difficult for non-linear projections \cite{gascuel2008fast,brosz2007single}.

We see that rolling images contain the expected non-linear projection effects: long edges that are straight in 3D appear as curves in the image.
As this mapping is consistent, other effects such as shadows and specularities appear consistent for all approaches.
Warping however has difficulties with disocclusions, edges and fine details.
We see that large parts of the background are missing.
The biggest challenge are areas occluded in the input image.
Large parts are missing in warping, \eg the sky background in \textsc{Helicopter} condition, and the ground plane in \textsc{Houses}, that are easily resolved by our approach.
Current Warping techniques always have difficulties with edges, where a pixel can only be either warped or not, resulting in jagging artifacts such as on the edges of \textsc{Children}.
When motion, occlusion and fine edge structures come together, such as in the area around the \textsc{Helicopter}'s rotor, the warped images bear little resemblance to the reference.

\myparagraph{Joint rasterization}
Results for joint rolling-foveated images are show in \refFig{JointRasterization}.
We see both the expected improvement in the foveal inset and the global rolling: the car and fence have straight 3D edges that turn into curves under viewer motion.
Those scenes have around 100,00\,k faces and render in less than 50\,ms.

\mycfigure{JointRasterization}
{
Joint \ie rolling and foveated, perceptual rasterization for three scenes.
The insets compare joint and rolling-only results.
}

\myparagraph{Lens Distortion}
Including a barrel lens distortion \cite{OculusSDK} in the joint approach is simple (\refFig{LensDistortion}): we just use a foveation function $p$ that is a composition $p(d)=p_\mathrm c\circ p_\mathrm l(d)$ of the cortical foveation $p_\mathrm c$ function and a lens distortion function $p_\mathrm l$.
When sampling back from the foveated domain, only $p_\mathrm c$ is applied, as $p_\mathrm l$ will happen optically.
Only the -- much smaller -- chromatic blur still needs to be applied, as the effort of rasterizing three channels independently does not appear justified.

\myfigure{LensDistortion}{This stereo image is both rolling and foveated, as well as it will appear lens-undistorted in space and chroma when observed in a \ac{hmd}.}

\myparagraph{Rolling Shading}
Here we compare rolling shading, included in all the above results, to rolling rasterization without rolling shading in \refFig{Specular}.
Specular inconsistencies will result in popping artifacts over time \cite{antonow2015}, where a highlight does not slide across the side of the car but judders between frames.

\myfigure{Specular}{
Rolling rasterization without rolling shading \emph{(left)} lack some specular effects.
Rolling shading \emph{(right)} produces highlights that change across the image due to the change in view over time.
}

\mysubsection{Quantitative}{QuantitativeResults}
Here alternatives and variants of our approach are compared in terms of speed and image similarity.

\myparagraph{Methods}
We tested our approach on a Nvidia Quadro~K6000.
Image similarity is measured in terms of an adapted SSIM \cite{wang2004image} metric.
It ignores all disoccluded pixels, \ie it provides an upper bound on quality to what any hole filling, however sophisticated, could do \cite{didyk2010perceptually,schollmeyer2017efficient}.
For foveated comparisons, SSIM is computed for the $64\times 64$ foveal pixels.
For foveation, we compare speed and quality to a common method, that directly operates at the same resolution as ours, and speed to a three-layered method \cite{guenter2012foveated} assuming it will provide similar quality.
The foveated reference is a $8\times8$ super-sampled rasterized image.
Shading effort (SSAO and IBL) is the same for ours and common, while it is three times larger for layered and 16 times larger for the reference. 
As rolling methods we compare ``No rolling'' corresponding to the first column in \refFig{RollingResults}, ``Warping'' from the second column in \refFig{RollingResults} and our ``Rolling'' approach from the third column in \refFig{RollingResults}.
The rolling reference is ray-traced, that is, identical to all images perceptual rasterization produces.
We state the ray-tracing time of a reasonably implemented GPU traversal of an SAH-optimized BVH.

\myparagraph{Comparison}
Foveation results are shown in (\refTbl{FoveatedMain}).
Our approach is more similar to the reference than common rasterization.
Furthermore, it achieves speed that is roughly half as fast rasterizing multiple layers and very similar to rendering in a full resolution.
Finally, we see that refined bounds increase sample test efficiency as well as actual compute time.

Rolling results are stated in \refTbl{RollingMain}. 
First, we see that rolling and non-rolling images are substantially different according to the SSIM metric.
At the same time, classic GPU rasterization is highly optimized and produces images very quickly.
When warping the image, the similarity increases, but time is increased by two milliseconds: high-quality warping requires two primitives per pixel \cite{mark1997post}.
Next, we compare our method using different bounds.
Note, that the SSIM is always 1 as our rasterization has been verified to be identical to ray-tracing a rolling shutter image.
We also note, that scenes with many polygons, such as \textsc{Children} (1.4\,M) are feasible, but noticeably slower, likely due to the straightforward convex hull implementation used in the GP.

For both foveation and rolling, ray-tracing -- while very convenient and clean to implement -- is slower than all versions of rasterization.
Note, that the ray-tracing numbers do not include the SAH building time required, which is likely substantially larger.

Overall, perceptual rasterization achieves quality similar to a reference, while being slower than highly-optimized, fixed-pipeline rasterization by a moderate factor, but much faster than ray-tracing.

\begin{table*}
\centering
\setlength{\tabcolsep}{2.1pt}
\caption{
   Qualitative evaluation of foveated rasterization from \protect\refFig{FoveatedResults}.
   Layered SSIM is assumed to be 0.
   }%
   \vspace{-.25cm}
   \begin{tabular}{l r r r rr rr rr rrr rr r rr}
   \toprule
&
Tris&
\multicolumn{8}{c}{Ours}&
\multicolumn{3}{c}{Common}&
\multicolumn{2}{c}{Layered}&
Raytrace&
\multicolumn{2}{c}{Reference}
\\
&&
\multicolumn{2}{c}{}&
\multicolumn{2}{c}{Trivial}&
\multicolumn{2}{c}{Quad}&
\multicolumn{2}{c}{Recursive}\\
&&
Sim.&Shade&
Raster&STE&
Raster&STE&
Raster&STE&
Sim.&Shade&Raster&
Shade&Raster&
Raster&
Shade&Raster\\
\midrule
\textsc{Hairball}&
115\,k&
.988& 2.2\,ms&
28.7\,ms & 5.0\,\%&
 9.8\,ms&10.1\,\%&
 5.2\,ms&40.0\,\%&
.970& 2.2\,ms & 1.1\,ms& 
 6.6\,ms&2.7\,ms& 
280.0\,ms&
228.4\,ms &3.5\,ms\\
\textsc{Car}&
178\,k&
.992& 2.2\,ms&
35.2\,ms&  1.0\,\%&
12.2\,ms& 14.3\,\%&
 6.5\,ms& 37.3\,\%&
.962& 2.2\,ms& 1.4\,ms&
 6.6\,ms & 3.3\,ms&
33.2\,ms&
228.4\,ms&4.6\,ms\\
\textsc{Children}&
1,400\,k&
.992& 2.2\,ms&
3\,s&  0.0\%&
30.0\,ms& 16.0\,\%&
29.3\,ms& 48.2\,\%&
.938& 2.2\,ms& 2.9\,ms&
  6.6\,ms& 7.9\,ms&
48.5\,ms&
228.4\,ms &13.8\,ms\\
\bottomrule
\end{tabular}
\label{tbl:FoveatedMain}
\end{table*}

\begin{table*}[ht]
	\centering
	\setlength{\tabcolsep}{3.5pt}
	\caption{
   Qualitative evaluation of rolling rasterization from \protect\refFig{RollingResults}.
   Our SSIM is 1 in all conditions.
   }%
   \vspace{-0.25cm}
	\begin{tabular}{l r rr rr rrrrrrrrr r}
    \toprule
    Scene&
    Tris&
    \multicolumn{2}{c}{No Rolling}&
    \multicolumn{2}{c}{Warping}&
    \multicolumn{9}{c}{Rolling (ours)}&
    Raytrace
    \\
    &
    &
    &
    \multicolumn{2}{c}{}&
    \multicolumn{2}{c}{}&
    \multicolumn{2}{c}{Quad}&
    \multicolumn{2}{c}{Hull}&
    \multicolumn{2}{c}{Adaptive}&
    \multicolumn{2}{c}{Zenon}
     \\
    &
    &
    Sim. & Time &
    Sim. & Time &
    Sim. &
    Time & STE &
    Time & STE &
    Time & STE &
    Time & STE & Time
    \\
    \midrule
    \textsc{Helicopter} &
    15\,k&
    .768 & 0.9\,ms &
    .700 & 2.5\,ms &
    1.00 &
    31.2\,ms & 4.2\%& 
    15.5\,ms & 9.7\%& 
    5.8\,ms & 36.9\%& 
    4.1\,ms & 48.1\%&
    30.4\,ms
    \\
    \textsc{Sponza} &
    223\,k&
    .177 & 1.9\,ms &
    .322 & 4.5\,ms &
    1.00 &
    3\,s & 0.0\%&
    361.9\,ms & 1.8\%&
    134.0\,ms & 5.5\%&
    38.5\,ms & 18.5\%&
    113.5\,ms
    \\
    \textsc{Houses} &
    13\,k&
    .674 & 0.7\,ms &
    .727 & 2.5\,ms &
    1.00 &
    25.0\,ms & 5.6\%&
    13.9\,ms & 12.2\%&
    6.7\,ms & 28.2\%&
    5.2\,ms & 39.1\%&
    16.0\,ms
    \\
    \textsc{Children} &
    1,400\,k&
    .610 & 3.8\,ms &
    .780 & 6.3\,ms &
    1.00 &
    80.0\,ms & 1.0\%&
    65.5\,ms & 2.2\%&
    36.7\,ms & 26.7\%&
    28.1\,ms & 37.2\%&
    45.3\,ms
    \\
	\bottomrule
\end{tabular}
\label{tbl:RollingMain}
\end{table*}

\myfigure{Plots}{
Comparison of different rolling approaches in \textsc{Helicopter}: Classic rasterization, warping and rolling.
\textbf{a)} Image resolution and compute time (less is better).
\textbf{b)} Transformation (a camera rotation) and compute time (less is better).
\textbf{c)} Transformation and image similarity in SSIM (more is better).
Scalability of foveation in \textsc{Car}: Similarity \emph{(yellow line)} and compute time \emph{(pink line)} as a function of foveation.
}

\myparagraph{Sample Test Efficiency}
We also compute the sample test efficiency (STE) \cite{mcguire2010real,akenine2007stochastic,laine2011clipless}, defined as the ratio of pixels belonging to a primitive to the number of pixels tested.
An STE of 100\,\% would mean that only necessary test were made, \ie the bounds were very tight.
A low STE indicates that unnecessary tests occurred.
Comparing the bounding approaches in \refTbl{FoveatedMain} and \refTbl{RollingMain}, it can be seen that investing computational effort into tight bounds, pays off with a higher STE and is ultimately faster overall.
Visualizations of the STE for rolling rasterization are seen in \refFig{STERolling}.

\myfigure{STERolling}{Sample test efficiency of different rolling bounds in \textsc{Helicopter}.
We see, that convex hulls are tighter than quads, but only bounds that adapt to the space-time structure have a workable  STE, where Zenon's is more tight to the right of the image where motion is largest.}

\myparagraph{Scalability}
Dependency of speed and image similarity on external variables is plotted for different approaches in \refFig{Plots}.
The first plot shows how image resolution affects computation time (\refFig{Plots}, a).

We see that our approach is, as expected, slower than common rasterization, which is highly-optimized in GPUs.
At the same time warping does not scale well with resolution due to the many pixel-sized triangles to draw. At high resolutions, the warping method is worse both in terms of speed, as well as image quality.

Next, we analyze computation time as a factor of the transformation occurring during the scan-out (\refFig{Plots}, b). We quantify this as view rotation angle around the vertical axis.
We see that classic rasterization is not affected by transformation at all.
Warping adds an almost-constant time overhead that only increases as larger polygons are to be drawn.
Our approach is linearly dependent. The amount of pixel motion is expected to be linear in small angles.
Our tighter bounds can at best reduce the magnitude of the linear relationship.
For large motions our approach is approximately half as fast as fixed-function rasterization plus warping, or six times slower than fixed-function rasterization alone. 

Next, we analyze similarity (more is better) depending on the transformation, again parametrized as an angle (\refFig{Plots}, c).
We find that our approach, as expected, has no error relative to the ray-tracing reference.
With no user motion, common rasterization has no error either, while warping still introduces pixel-sampling problems.
As motion becomes more extreme warping reduces error with respect to common rasterization, but similarity still decreases, as disocclusions cannot be resolved from a single image.

Finally, we see the dependency of similarity and compute time on foveation strength $\alpha$ (\refFig{Plots}, d), in the power foveation function $p(d)=d^\alpha$.
We find that similarity is a convex function, peaking around the value $\alpha=2$ we use.
Too low-a foveation does not magnify enough to benefit from the super-sampling. Too high values magnify so much, that only the central part of the fovea benefits, reducing SSIM again.
Time is a linear function of foveation strength, as polygonal bounds to increasingly curved triangles are decreasingly tight.

\myparagraph{Head Pose Estimation}
Finally, we investigate the effect of head pose prediction error on our approach.
Before (\eg \refFig{Plots}, c), we have seen that the image error is proportional to the error in transformation.
Therefore, we sampled head motion using the DK2 at approximately 1000\,Hz. At each time step we used the SDK's predictor - the same that drives the rolling rasterization - to predict the pose one frame ahead. We use these captures to determine how the linearly interpolated pose and a time-constant pose differ from the actual pose. For 3,459 frames of typical DK2 motion, we we found the \textsc{linear} prediction to have an error of .001~meter in translation and .25~degree in rotation while the error of a \textsc{constant} prediction is much larger, at .05~meter and 1.3~degrees, indicating a linear model already removes most of the overall error.

\mysection{Perceptual Evaluation}{UserStudies}
To quantify the perceptual effect of our technique, we conducted four user studies:
a threshold estimation experiment to establish the optimal foveation for a particular apparatus (\refSec{FoveationStudyStaircase});
an image judgment experiment comparing super-sampling in our foveation approach to a reference super-sampled image (\refSec{AntiAliasingStudy});
an object tracking experiment with and without rolling rasterization (\refSec{RollingTaskStudy}) and
an image preference experiment comparing our rolling approach to other approaches such as warping in an \ac{hmd} (\refSec{RollingStudyPreference}).

\mysubsection{Foveation Strength}{FoveationStudyStaircase}
This study demonstrated that there was no perceptual difference between a non-trivially foveated image and a traditionally rendered image.
We based our protocol on that of Patney~\etal~\shortcite{patney2016towards}, performing a 2AFC staircase task to identify the \ac{jnd} threshold - the foveation strength at which participants begin to reliably detect foveation artefacts. 

\myfigure{FoveationStudy}
{
Foveation study stimuli \textbf{(a)} and analysis (Please see text) \textbf{(b)}.
}

\myparagraph{Procedure}
After being fitted with an eye-tracker, participants engaged in a 2AFC task.
In each trial, participants were exposed to two 1.5 second sequences of a rotating model - one with foveation and one traditionally rendered - with a .75-second gap in-between. The rotation was around the vertical axis at one revolution every 14~seconds.
After viewing both sequences, participants were asked to indicate via the keyboard which of the two was ``higher quality".
The order of rendering technique was randomized.
Participants each completed 180 trials on one of three models.
Foveation strength was determined by a 1-up/3-down staircase following the guidelines of Garcia-Perez~\&~Alcala-Quintana~\shortcite{Garcia-Perez2007}. 

\myparagraph{Apparatus}
We used a typical desktop PC with a GTX~980~GPU and an Asus~VG248 144\,Hz monitor to render the scenes with image-based lighting and specular materials (\refFig{FoveationStudy},a) under natural HDR illumination. The eye-tracker was an SR-Research EyeLink~II connected via Ethernet directly to our application. 

\myparagraph{Participants}
25 na\"ive participants successfully completed our study across three conditions: \textsc{Lucy} (7), \textsc{Rockbox} (9), \textsc{CAD} (9). 

\myparagraph{Analysis}
We opted for a fixed-size staircase with empirically set step-sizes, as our technique is novel and we do not have any reasonable priors for parametric sampling schemes.
For our analysis though we fit a logistic psychometric function \cite{Garcia-Perez2007} for simplicity and comparability 
to estimate thresholds and confidence intervals at 95\,\%.

\myparagraph{Results}
\refFig{FoveationStudy} shows an approximate psychometric function computed by averaging the function parameters for each participant, for each condition.
A psychometric function describes the probability of detecting a distortion (vertical axis) depending on the foveation strength (horizontal).
We see that the 75\,\% detection probability JND threshold occurs at non-zero levels of foveation.
This indicates subjects cannot detect our foveation even when present at such strengths.
The confidence intervals (colored bars) show the significance of this observation.
Participant's individual functions, staircase results and further analysis are included in our supplementary materials.


\mysubsection{Super-sampled Image Comparison}{AntiAliasingStudy}
We performed a second study to determine how users compare our foveally super-sampled images to traditionally rendered images, when presented with a super-sampled image as a reference.

\myparagraph{Protocol}
After being fitted with an eye-tracker, participants engaged in a series of \ac{2afc} trials.
In each trial, participants viewed three instances of a slowly rotating model side-by-side (\refFig{FoveationStudy}). 
The center model was a 4$\times$4 super-sampled reference, with common and foveated to the sides in a randomized order.
Participants were asked to indicate via the keyboard which side appeared most similar to the reference.
Subjects each completed 45 trials spread evenly across three conditions, randomly interleaved.

\myparagraph{Participants}
7 na\"ive participants completed this study. 

\myparagraph{Apparatus}
We used the same apparatus as the previous study.

\myparagraph{Analysis}
We compute the preference for foveation as the proportion of aggregated trials in which the foveated image was chosen, for each condition.
Two-tailed binomial tests ($n=105$) indicated that the preferences are significantly different to chance ($p=.5$) for all conditions ($p < .001$);

\myparagraph{Results}
The results show a strong and consistent preference for foveation across all models (\textsc{\textoverline{Lucy}}: 90\%, \textsc{\textoverline{Flower}}: 94\%, \textsc{\textoverline{CAD}}: 87\%). 
The slight reduction for CAD is likely because when viewing the back of the model there were few details to distinguish the techniques.

\mysubsection{Rolling Rasterization Task Performance}{RollingTaskStudy}
We conducted a user study to examine how rolling rasterization affects perception in \ac{vr}.
We used an object tracking task to measure how behavior is influenced by both rolling rasterization and the asynchronous time-warping.

\myparagraph{Protocol}
Participants were shown a simple virtual environment in which a \SI{50}{\centi\metre} box moved along a \ang{180} curve, \SI{8}{\metre} in front of them, just above eye level.
The box reversed direction at the extents and moved at \sipm{85.9}{68.7}{\degree\per\second}, the rate changing randomly every second.
A head-fixed reticle was visible \SI{8}{\metre} ahead.

\myfigure{Stimuli}{Stimuli of the first \textbf{(a)} and second \textbf{(b)} experiment.
\textbf{c)} Probability Functions of Phase for each condition.
}

Participants were told to use their head to keep the reticle in the middle of the box. 
Participants followed the box in two trials, each lasting 4 minutes.
Between the trials participants had a short break outside the \ac{hmd}.
There were three conditions pertaining to the rasterization method used: traditional (STD), Oculus' Asynchronous Time-warping (ATW) and our Rolling-Rasterization (ROL).
The conditions were presented in 30-second-blocks, randomly interleaved.

\paragraph{Participants} 
20 na\"ive participants completed the study. 


\myparagraph{Apparatus}
Our experiment was performed with an Oculus Rift DK2.
This HMD has a single low-persistence rolling-scanout display that scans right-to-left at 75~Hz with a persistence of \SI{4}{\milli\second} (\refFig{Capture}).
The DK2 has an inertial measurement unit (IMU) that samples at \SI{1}{\kilo\hertz}.
The head and box positions were sampled at \SI{75}{\hertz}.

\myparagraph{Analysis and Results}
We began by analyzing the phase of the head motion.
This is the instantaneous angular difference between the head and the box, positive when the head is leading the box, and negative when it is following.
If participants were tracking the box exactly, we would expect a symmetrical distribution with a slight negative bias due to latency.
Instead, a set of Kolmogorov-Smirnov tests show a non-normal distribution for the conditions separately and cumulatively ($P<0.05$).
All conditions also show a positive bias (\refFig{Stimuli}, c).
This indicates a tendency to lead the target. 


\begin{table}[htbp]
  \centering
  \caption{ANOVA test results for mixed-model terms}%
  	\vspace{-.25cm}
    \begin{tabular}{p{2.2cm} p{0.8cm} r p{0.3cm} p{0.7cm} l}
    \toprule
    Variable & Effect & \multicolumn{1}{l}{F-Stat} & \multicolumn{1}{l}{DoF} & \multicolumn{1}{l}{DoF} & \multicolumn{1}{l}{$p$} \\
    \midrule
    Per-Parti.\ Inter.\ & Random & 1394.14 & 1 & 501,799 & $<$.0001 \\
    Condition & Fixed & 114.53 & 2 & 501,799 & $<$.0001 \\
    Speed & Fixed & 7984.79 & 1 & 501,799 & $<$.0001 \\
    Condition: Speed & - & 134.06 & 2 & 501,799 & $<$.0001 \\
    \bottomrule
    \end{tabular}%
  \label{tbl:study_resultstable}
\end{table}%

In this case, we would expect the lead to increase as apparent latency decreases, and this is what is shown.
ROL enables the largest anticipatory behavior (a lead of \SI[parse-numbers=false]{\overline{2.7}}{\degree}).
STD presents the second largest (\SI[parse-numbers=false]{\overline{2.2}}{\degree}), having a latent but head-fixed image.
ATW has the smallest lead (\SI[parse-numbers=false]{\overline{1.7}}{\degree}), because while it compensates for latency, it does so by moving the entire image - including the target - counter to head rotation introducing apparent lag into the target.

To test the significance of this, we performed an ANOVA on the terms of a linear-mixed model, to control for per-participant biases and speed.
The results of this test are shown in \refTbl{study_resultstable}, indicating a highly significant effect of rendering condition ($p<.0001$), as well as an interaction between rendering condition and speed ($p<.0001$).


\mysubsection{Qualitative Rolling Study}{RollingStudyPreference}

This study compares fidelity of rolling rasterization to a traditional warping technique using a set of two-alternative forced choice trials.
In this study the stimuli were head-fixed and in mono, to avoid any effects of latency or head motion.

\myparagraph{Protocol}
Participants were provided with the same HMD as before (\refSec{RollingTaskStudy}) and exposed to a set of video pairs.
Each video pair was seen consecutively and preference indicated using a keyboard.

\myparagraph{Stimuli}
The videos all showed  moving objects (\refFig{Stimuli}).
The videos were one of three types.
\textsc{Std} rendered each frame at a fixed time. 
\textsc{Warp} took each frame from STD and warped it from $t_\mathrm s$ to $t_\mathrm e$ with a traditional warping algorithm (Mark~et~al.'s~\shortcite{mark1997post} pixel-sized grid warping, where triangles stretching by more than a threshold are culled).
ROL rendered an image for $t_\mathrm s$ to $t_\mathrm e$ with our technique.
Each video was presented for 2.5~seconds.
Both the pairings of the types, and the order in which they were presented within the pair, were balanced an random.

\myparagraph{Participants}
9~na\"ive participants completed the study seeing all combinations five times. 

\myparagraph{Analysis and Results}
We summed the number of times each condition was preferred for each participant, and performed a one-way ANOVA on these three groups. We show a significant effect of rendering technique [$F(2,24)=13.45,p=.0001$], with users preferring STD or ROL over WARP. 
There is no significant difference between STD or ROL ($p=.993$).
These results indicate that rolling rasterization  can be indistinguishable from a traditional render in this protocol.

\mysection{Discussion}{Discussion}

\myparagraph{Comparison to warping}
\myfigure{Protocoll}{
Conceptual differences of our approach and warping to ground truth when producing frame $n+1$.
Time is the horizontal axis and the vertical axis is view transformation.
Differences in images, and by this the perceived error, are likely proportional to differences in view transform (dotted lines).
The ground truth view transform is shown as a single function curve for frame $n$ and $n+1$.
Different methods are encoded as colors.
Colored horizontal blocks are working time, colored lines are approximations of the view transform for images on display.
}
The differences between an ideal ground truth zero-latency rasterizer, warping and our approach is seen in \refFig{Protocoll}.
The ground truth approach (orange), would instantaneously produce an image that at every position in space-time will match the view transform.
Normal rasterization preceding warping (light blue) will render frame $n+1$ with the transform known at time $t_1$.
By $t_3$, the end of frame $n+1$, the display image will be severely outdated (difference E1).
Warping (dark blue), will move pixels to compensate for the transformation at $t_2$, but can still not mitigate the image to become outdated during $n+1$, (difference E2) and it has no way to remedy disocclusions occurring between $t_1$ and $t_2$.
Our approach (green) also starts work at $t_1$, but using the transformation predicted for continuous points in time on frame $n+1$, removing all occlusion and shading error and leaving only the prediction error E3.
Even when assuming a hypothetical and unpublished competitor that rasterizes using a predicted view transform (dotted light blue line) and a rolling form of warping (dark blue dotted line), there remains an appearance error E4 at $t_4$ that can not ever be resolved by rasterizing outdated (\ie non-rolling) occlusion and shading.

\myparagraph{Fast Rendering}
It is tempting to just hope faster rendering will make rolling rasterization obsolete.
But any common non-rolling method will never reduce latency below the scan-our duration, typically around 16 \,ms.
Even if a fast non-rolling rasterization takes only 1\,ms (a short light-blue bar in \refFig{Protocoll}), the scan-out still takes 16\,ms, and the latency will remain to be 15\,ms.
Using rolling rasterization, that might be slower, say 4\,ms, (green bar longer than the light-blue bar in \refFig{Protocoll}) would be better deal, as the latency can get arbitrarily small if a sufficiently correct prediction is made. 

\myparagraph{Prediction}
Like any method that has to finish before the scan-out starts, we require a prediction of scene and viewer motion during the scan-out.
Grossmann~\etal\shortcite{grossman1988frequency} have measured the velocity and acceleration of head motions.
Their results show that rotational and translational head velocity can be substantial, indicating, that the rendering with a view transform that changes during the display interval is useful.
They also find, that the acceleration \ie derivation form a linear model, is small as it requires force.
This indicates that our first order-model, with substantial velocity but limited acceleration, is physiologically plausible.

\myparagraph{Streaming}
Friston~\etal\shortcite{friston2016construction} update the view matrix for each scan-line and ray-trace a simplistic scene in a period far below that of the display's scan out.
It would not be clear how to ray-trace a complex scene in this time.
Geometry in animated scenes changes for every scan line, which would require a very high frequency of BVH rebuilds when using ray-tracing.
In our case of streaming OpenGL rasterization, which maps primitives to pixels, we have no guarantees on the space or time layout of the primitive stream.
Consequently, we need to predict the head pose across the scan-out.
Prediction is essential and cannot be omitted.
Even if a sensor could give the absolute viewpoint continuously, there is still the delay due to rendering the image from this viewpoint, and therefore an interval between the rasterization and the actual scan-out.
We further assume the change in transformation is small enough that the transform matrices can be linearly interpolated; an optimization that could be replaced with a more advanced interpolation.

\myparagraph{Speed}
We demonstrate a prototypical implementation using a GPU, which has speed comparable non-rolling or non-foveated implementation.
Our current implementation runs at real-time rates, suggesting a full hardware implementation (with optimizations such as tiling, etc. \cite{akenine2007stochastic}) could achieve speeds similar to a traditional rasterizer.

\myparagraph{Joint analysis}
We have derived bounds for joint foveated-rolling rasterization and show example results in \refFig{JointRasterization}, but did not conduct a perceptual (stereo) experiment for this combination.

\myparagraph{Periphery}
Similar to other methods \cite{guenter2012foveated,patney2016towards,stengel2016adaptive} our foveated rasterization can create temporal aliasing in the periphery, where humans are unfortunately particularly sensitive.
Future work will investigate specialized spatio-temporal filters to circumvent this issue. 

\myparagraph{Screen-space effects}
Screen space shading needs to be adapted to support perceptual rasterization.
We have done so for SSAO by multiplying all image distances by the pixel density $p(x)$.

\mysection{Conclusion}{Conclusion}
In this paper we introduced a new efficient rasterization technique that exploits the spatio-temporal-retinal relation of rays and primitives found in \acp{hmd}.
It prevents the artifacts and overhead of warping and works in a single pass while supporting moving objects, viewer translation and rotation as well as specular shading and lens distortion - all of which are challenging for warping.
The main technical contribution is the derivation of tight and efficiently computable pixel-primitive bounds.

Future investigations could extend the rolling concept to physics and other simulations, and would also need to seek better understanding of the relationship between latency and motion blur, focus and the role of eye and head motion.
We only touched upon the relation to stereo or even light field displays.


\bibliographystyle{ACM-Reference-Format}
\bibliography{article}

\end{document}